\pdfoutput=1
\documentclass[aip,amsmath,amssymb, reprint]{revtex4-1}

\usepackage{prelim2e} 

\usepackage{graphicx}
\usepackage{dcolumn}
\usepackage{bm}
\usepackage{booktabs}
\usepackage{multirow}
\usepackage[utf8]{inputenc}
\usepackage[T1]{fontenc}
\usepackage[version=3]{mhchem}
\usepackage{hyperref} 
\usepackage{rotating} 

\newcommand{\plh}{%
   \ensuremath{\times}
}

\usepackage{notes2bib}

\newcommand{\scinumberEh}[2]{\ensuremath{#1 \times 10^{#2} E_{\text{h}}}}

\newcommand*{\citeref}[1]{Ref.~\citenum{#1}}

\usepackage{refstyle}
\newref{subsec}{
  name = Subsection~,
  names = Subsections~,
  Name = Subsection~,
  Names = Subsections~
}

\usepackage{braket}
\usepackage[table]{xcolor}
\usepackage{makecell}
\usepackage{listings}
\usepackage{bbold}
\usepackage{url}
\usepackage{mathtools}

\usepackage{colortbl}

\usepackage{placeins}
\usepackage{subfigure}
\usepackage{changes}

\definecolor{alekseicolour}{RGB}{153, 76, 0}

\definechangesauthor[name={Aleksei}, color={alekseicolour}]{ai}

\newcommand{\PySCF}{\textsc{PySCF}}

\definecolor{darkolivegreen}{rgb}{0.33, 0.42, 0.18}
\definecolor{darkpink}{rgb}{0.91, 0.33, 0.5}
\definecolor{aqua}{rgb}{0.0, 1.0, 1.0}
\definecolor{applegreen}{rgb}{0.55, 0.71, 0.0}
\definecolor{blizzardblue}{rgb}{0.67, 0.9, 0.93}
\definecolor{bleudefrance}{rgb}{0.19, 0.55, 0.91}
\definecolor{desertsand}{rgb}{0.93, 0.79, 0.69}
\definecolor{mediumspringbud}{rgb}{0.79, 0.86, 0.54}
\definecolor{rosequartz}{rgb}{0.67, 0.6, 0.66}

\definecolor{armygreen}{rgb}{0.29, 0.33, 0.13}
\definecolor{asparagus}{rgb}{0.53, 0.66, 0.42}
\definecolor{aquamarine}{rgb}{0.5, 1.0, 0.83}
\definecolor{cambridgeblue}{rgb}{0.64, 0.76, 0.68}

\definecolor{background-color}{gray}{0.98}

\usepackage{soul}

\begin{document}

\title{How good are recent density functionals for ground and excited states of one-electron systems?}
\author{Sebastian Schwalbe}
\email{schwalbe@physik.tu-freiberg.de}
\affiliation{\mbox{Institute of Theoretical Physics, TU Bergakademie Freiberg, D-09599 Freiberg, Germany}}
\author{Kai Trepte}
\email{kai.trepte1987@gmail.com}
\affiliation{\mbox{Taiwan Semiconductor Manufacturing Company North America, San Jose, CA 95134, USA}}
\author{Susi Lehtola}
\email{susi.lehtola@alumni.helsinki.fi}
\affiliation{Molecular Sciences Software Institute, Blacksburg, VA 24061, USA}

\date{\today}

\begin{abstract} 
Sun et al. [J. Chem. Phys. 144, 191101 (2016)] suggested that common
density functional approximations (DFAs) should exhibit large energy
errors for excited states as a necessary consequence of orbital
nodality. Motivated by self-interaction corrected density functional
calculations on many-electron systems, we continue their study with
the exactly solvable $1s$, $2p$, and $3d$ states of 36 hydrogenic
one-electron ions (H--\ce{Kr^{35+}}) and demonstrate with
self-consistent calculations that state-of-the-art DFAs indeed exhibit
large errors for the $2p$ and $3d$ excited states. We consider 56
functionals at the local density approximation (LDA), generalized
gradient approximation (GGA) as well as meta-GGA levels, also
including several hybrid functionals like the recently proposed
machine-learned DM21 local hybrid functional. The best non-hybrid
functional for the $1s$ ground state is revTPSS. The $2p$ and $3d$
excited states are more difficult for DFAs as Sun et al. predicted, and
LDA functionals turn out to yield the most systematic accuracy for
these states amongst non-hybrid functionals. The best performance for
the three states overall is observed with the BHandH global hybrid GGA
functional, which contains 50\% Hartree--Fock exchange and 50\% LDA
exchange. The performance of DM21 is found to be inconsistent,
yielding good accuracy for some states and systems and poor accuracy
for others. Based on these results, we recommend including a variety
of one-electron cations in future training of machine-learned density
functionals.
\end{abstract}

\maketitle

\hyphenation{work-horses}

\section{Introduction \label{sec:intro}}

Density-functional theory \cite{Hohenberg1964_PR_864,
Kohn1965_PR_1133} (DFT) has become one of the workhorses of
computational chemistry, material science, and related fields, as
modern density functional approximations (DFAs) only require a
reasonable amount of computational effort while providing a level of
accuracy sufficient for semi-quantitative predictions on a broad range
of systems.\cite{Becke2014_JCP_18, Jones2015_RMP_897,
Mardirossian2017_MP_2315} Although hundreds of DFAs have been
proposed, thus forming the infamous zoo of
DFAs,\cite{Goerigk2019_AJC_563} new DFAs continue to be developed in
the aim to find more universally applicable DFAs that combine suitable
levels of accuracy and numerical effort.

New DFAs can be constructed along various
strategies.\cite{Barth2004_PS_9, Cohen2012_CR_289,
Mardirossian2017_MP_2315, Lehtola2018_S_1} The traditional route to
construct DFAs is to start from first principles and to impose known
limits and constraints; this is the way along which many well-known
functionals such as PBE\cite{Perdew1996_3865},
TPSS\cite{Tao2003_146401, Perdew2004_6898}, and
SCAN\cite{Sun2015_036402} have been constructed.

Semi-empirical fitting is another route for constructing DFAs. Here,
the general idea is to introduce flexibility in the functional form by
introducing several independent DFA components that are weighted by
parameters which are optimized against some training
dataset. Classical examples of semi-empirically fitted functionals
include B3LYP,\cite{Stephens1994_11623} Becke's 1997
functional\cite{Becke1997_JCP_8554} (B97) and several refinements
thereof such as the HCTH [Hamprecht--Cohen--Tozer--Handy] functionals
by Handy and coworkers,\cite{Hamprecht1998_6264,
Boese2000_1670} as well as the Minnesota family of DFAs by Truhlar
and coworkers\cite{Zhao2005_161103, Zhao2006_364} that has
been reviewed by \citet{Mardirossian2016_JCTC_4303}.

In reality, the classification of functionals into ones built solely
from first principles vs ones formed by semi-empirical fitting is not
always clear: for instance, the TPSS exchange functional is
parametrized to yield the exact energy for the hydrogen atom's exact
ground state density,\cite{Tao2003_146401, Perdew2004_6898}
while the SCAN functional\cite{Sun2015_036402} includes parameters
that are fit to data on noble gases. Modern semi-empirical
functionals,\cite{Mardirossian2014_PCCP_9904,
Mardirossian2015_074111, Mardirossian2016_214110,
Brown2021_2004, Sparrow2022_JPCL_6896, Trepte2022_1104} in
turn, typically employ a combination of the two approaches by
restricting the fits to known constraints.

DFAs from either route are widely used, given their suitable numerical
accuracy and reasonable computational effort.  However, the
functionals obtained from the two routes tend to exhibit different
behavior. For instance, while semi-empirical DFAs often deliver
excellent descriptions of the total energy, they may fail to reproduce
electronic densities of the same quality: a famous article
of \citet{Medvedev2017_S_49} initiated an intense debate about this in
the literature,\cite{Medvedev2017_S_49, Kepp2017_S_496,
Medvedev2017_S_496, Kepp2018_PCCP_7538, Su2018_PNASUSA_2287,
Wang2017_JCTC_6068}; it was even pointed out that any general
mathematical measure of density error is too arbitrary to be
universally useful.\cite{Sim2018_JPCL_6385} DFAs built on physical
first principles, in contrast, often yield steady performance in a
variety of applications, but may not achieve the same level of
accuracy as tailored functionals for specific types of systems.

One of the most important limitations of present-day DFAs, regardless
of their design, is the self-interaction error (SIE): an artificial
interaction of the electrons with themselves. This error is related to
density delocalization error and the fractional electron
problem,\cite{Cohen2010_, Bryenton2022_WCMS_1631} and leads to
incorrect dissociation limits\cite{Ruzsinszky2006_JCP_194112} and
barrier heights,\cite{Johnson1994_CPL_100} for instance.
Recent avenues for circumventing SIE in DFAs involve determining the
electron density with another method, such as
Hartree--Fock\cite{Kim2013_PRL_73003, Sim2022_JACS_6625} or
multiconfigurational wave function theory\cite{Bao2017_5616,
Bao2018_JPCL_2353}. Other types of approaches have also been proposed
in the literature. To solve the self-interaction
problem, \citet{Perdew1981_PRB_5048} (PZ) proposed an
orbital-by-orbital self-interaction correction (SIC)
\begin{equation}
    E^\text{PZ} = E^\text{KS} - \sum_{i\sigma} \Delta_{i\sigma}, \label{eq:pz}
\end{equation}
where $E^\text{KS}$ is the Kohn--Sham (KS) energy
functional\cite{Kohn1965_PR_1133} and the self-interaction error (SIE)
is defined by
\begin{equation}
\Delta_{i\sigma} = E_J[n_{i\sigma}]+E_\text{xc}[n_{i\sigma}]. \label{eq:sie}
\end{equation}
Here, $n_{i\sigma}$ is the electron density of the $i$-th occupied
orbital with spin $\sigma$ and $E_J$ and $E_\text{xc}$ denote the
Coulomb and exchange-correlation energy functionals, respectively.
The idea behind PZ-SIC is that the self-interaction error defined
by \eqref{sie} vanishes for the exact
functional,\cite{Perdew1981_PRB_5048} and thereby the Perdew--Zunger
functional of \eqref{pz} is a better estimate for the total energy
than the uncorrected Kohn--Sham DFA $E^\text{KS}$; indeed, the PZ
functional is exact for one-electron systems such as the \ce{H2+}
molecule with approximate DFAs.

Despite the simple logic used to construct the PZ-SIC functional, the
PZ-SIC method turns out to be quite complicated. The introduction of
the explicit orbital dependence in \eqref{pz, sie} breaks the unitary
invariance of the energy functional,\cite{Pederson1984_JCP_1972}
requiring costly unitary optimization of the orbitals
(see \citeref{Lehtola2014_JCTC_5324} for discussion). However, even
though the resulting method is known to correct charge transfer errors
and barrier heights, it does not lead to improved atomization energies
with GGA and meta-GGA functionals in general.\cite{Perdew2015__193}

Continued research has illuminated other important theoretical aspects
of PZ-SIC. First, the orbital dependence in \eqref{pz, sie} has been
recently shown to require the use of complex-valued orbitals for
proper minimization, as real-valued orbitals can be shown to
correspond to high-order saddle points.\cite{Lehtola2016_JCTC_3195}
When complex-valued orbitals are employed, the total energy is
lowered, and PZ-SIC does lead to improved atomization energies for
some GGA functionals; however, more accurate atomization energies can
be obtained at significantly smaller cost with several standard
DFAs.\cite{Lehtola2016_JCTC_4296}

Second, the orbital dependence in \eqref{pz, sie} has also been shown
to lead to the existence of several local minima in the orbital
space.\cite{Lehtola2016_JCTC_3195} This problem has been recently
shown to persist also in a related SIC
method\cite{Pederson2014_JCP_121103} based on the use of
Fermi--L\"{o}wdin orbitals (PZFLO-SIC), where various choices for the
orbital descriptors lead to distinct local electronic
minima.\cite{Trepte2021_JCP_224109} The existence of such local minima
is a significant and underappreciated aspect of PZ-SIC and PZFLO-SIC
calculations, as finding the true ground state may require extensive
sampling of the space of the various possible localized electronic
configurations or bonding situations.

Despite their theoretical shortcomings, PZ-SIC and PZFLO-SIC have been
found useful in many applications\cite{Cheng2016_NC_11013,
Zhang2016_JPCL_2068, Ivanov2021_JPCL_4240} and we are positive that
several of the aforementioned issues in PZ-SIC and PZFLO-SIC can be
addressed by developments in the related theories by changing the way
the self-interaction correction is applied. One possible way to
achieve improved results would be to revisit DFAs based on the
requirements of SIC calculations.\cite{Jonsson2015_PCS_1858} It is
known that present-day DFAs yield poor estimates for the noded
electron densities that are involved in SIC
calculations.\cite{Sun2016_JCP_191101,
Shahi2019_JCP_174102} \citet{Sun2016_JCP_191101} demonstrated that the
ground and excited state densities of the hydrogen atom (as well as
of \ce{H2+}, see below) lead to large relative errors in the
exchange-correlation energy compared to the exact values, but we are
not aware of any self-consistent calculations on this issue.

Following recent discussion in the literature on the accuracy of DFAs
on the electron densities of small atoms and
ions\cite{Medvedev2017_S_49, Kepp2017_S_496, Medvedev2017_S_496,
Kepp2018_PCCP_7538, Su2018_PNASUSA_2287, Sim2018_JPCL_6385,
Wang2017_JCTC_6068} and motivated by the obvious connection of
one-electron errors (OEEs) to the PZ-SIC and PZFLO-SIC methods, in
this work we will analyze the OEE of various functionals for the $1s$
ground state as well as the $2p$ and $3d$ excited states of hydrogenic
ions $Z^{(Z-1)+}$, whose exact energies are well-known to be given in
atomic units by
\begin{equation}
    E_n = -Z^2/2n^2, 
    \label{eq:Eexact}
\end{equation}
where $Z$ is the atomic number, $n \geq l+1$ is the principal quantum
number, and $l$ is the angular momentum.

As was mentioned above, calculations of ground and excited states of
the hydrogen atom and of the $1\sigma_g$ ground state and $1\sigma_u$
excited state of \ce{H2+} have been discussed
by \citet{Sun2016_JCP_191101} with non-self-consistent electron
densities, while the $1s$ ground states of hydrogenic mononuclear
cations as well as the $1\sigma$ ground states of hydrogenic diatomic
cations have been discussed recently
by \citet{Lonsdale2020_PCCP_15805} using self-consistent
calculations. The novel contribution of this work is to address
(highly) excited states with noded electron densities of hydrogenic
cations self-consistently.  Importantly, like the $1s$ ground state,
the $2p$ and $3d$ excited states (as well as the analogous higher
excited states like $4f$) are the lowest states of the corresponding
symmetry, and the ground-state Kohn--Sham scheme is applicable to such
excited states as well as shown by \citet{Gunnarsson1976_PRB_4274}.

We pursue thorough density functional investigations of the $1s$,
$2p$, and $3d$ states of hydrogenic ions in benchmark-quality Gaussian
basis sets specially suited for this purpose with a selection of 56
popular DFAs, including the recently developed, highly sophisticated
machine-learned DeepMind 21 (DM21) local hybrid
functional.\cite{Kirkpatrick2021_1385}


The layout of this work is as follows. The computational details are
presented in \secref{compdet}, and the results are given
in \secref{results}. A summary of our findings and an outlook for
further investigations is given in \secref{summary}. Atomic units are
used throughout, unless specified otherwise.

\section{Computational details \label{sec:compdet}} 

We only use free and open-source software (FOSS) in this work,
following the philosophy discussed in \citeref{Lehtola2022_WIRCMS_1610}.
\PySCF{}\cite{Sun2020_JCP_24109} is an electronic structure code for
all-electron calculations using Gaussian-type orbitals (GTOs). 
As we are targeting one-electron states of specific symmetry ($s$,
$p$, or $d$ states), following \citet{Gunnarsson1976_PRB_4274} we
truncate the basis set in all calculations to contain functions only
of the pursued symmetry: calculations on the $1s$/$2p$/$3d$ state only
include the basis functions of the corresponding symmetry ($s$, $p$,
or $d$ functions, respectively) from the chosen parent basis set. This
procedure has two important features: the $2p$ and $3d$ excited states
become the ground state in the reduced-basis calculation, and the
computational requirements are smaller since fewer integrals need to
be calculated in the reduced basis than in the original basis set.

The one-electron guess---which is exact for one-electron systems and
thereby is also expected to be accurate for calculations employing
DFAs as well---is used in all
calculations.\cite{Lehtola2019_JCTC_1593} To ensure that the SCF
procedure converges to the global minimum instead of a saddle point,
the following procedure was used. First, a regular SCF calculation was
performed with \PySCF{} with default settings; direct inversion in the
iterative subspace (DIIS) is used to accelerate these
calculations.\cite{Pulay1980_CPL_393, Pulay1982_JCC_556} Next,
convergence to saddle point solutions was checked: cases where the SCF
converged to a final energy higher than that of the initial guess were
restarted, with new calculations employing iterative diagonalization
with level shifting\cite{Saunders1973_IJQC_699} instead of DIIS to
converge to the ground state. All calculations reported in this work
are fully converged to a threshold of \scinumberEh{1}{-7}.

For the GTO basis sets, we use the family of hydrogenic Gaussian basis
sets\cite{Lehtola2020_JCP_134108} (HGBS-$n$) that have been designed
for high-accuracy calculations on atoms and small molecules.  A
special feature of the HGBS basis sets is that the basis for atomic
number $Z$ is determined by a universal even-tempered basis set for
the ions $Y^{(Y-1)+}$ for $Y \in [1, Z]$, whereas augmented hydrogenic
Gaussian basis sets (AHGBS-$n$) add further functions for describing
the $Z=0.5$ one-electron ion.\cite{Lehtola2020_JCP_134108} The
parameter $n$ controls the relative precision of the hydrogenic
Gaussian basis; (A)HGBS-$n$ reproduces the exact total energies of the
one-electron ions to an approximate relative accuracy of
$10^{-n}$.\cite{Lehtola2020_JCP_134108} The motivation of this
approach in \citeref{Lehtola2020_JCP_134108} was that a many-electron
atom experiences a screened nuclear charge that can be rewritten in
terms of a radially dependent effective charge
$Z^\text{eff}=Z^\text{eff}(r)$ with the asymptotic limits
$Z^\text{eff}(0)=Z$ and either $Z^\text{eff}(\infty)=Z_\infty$ with
the asymptotic limit $Z_\infty=0$ for Hartree--Fock and DFT or
$Z_\infty=1$ for the exact effective
potential.\cite{Lehtola2019_JCTC_1593}

Another feature of the HGBS basis sets is that the functions of
various angular symmetries are determined independently of each other,
which facilitates the formation of polarized counterparts of the basis
sets that are essential for studying molecules and excited states, as
additional shells are added to the basis like lego
blocks.\cite{Lehtola2020_JCP_134108} Following
\citeref{Lehtola2020_JCP_134108}, the basis set with $p \ge 1$
polarization shells and accuracy $n$ is
denoted \mbox{(A)HGBSP$p$-$n$}. The definition of polarization shells
varies by atom (see
\citeref{Lehtola2020_JCP_134108} for discussion); however, as we only
include the functions of the pursued symmetry in each calculation, the
choice of the polarization level of the (A)HGBSP$p$-$n$ basis set does
not matter as long as the original basis contains functions of the
highest targeted angular momentum for the targeted atom, that is, $d$
functions in this work.

For the reasons listed above, the hydrogenic Gaussian basis sets of
\citeref{Lehtola2020_JCP_134108} are ideally suited for the
present study---as will be demonstrated in
\secref{results} by benchmarks with
functions from the polarization consistent (pc-$n$) basis
sets\cite{Jensen2001_JCP_9113} and their augmented
versions\cite{Jensen2002_JCP_9234} (aug-pc-$n$)---and, as will be
discussed in \secref{basconv}, we will take the exponents from the
AHGBSP3-$n$ basis sets in this work.  All basis sets were taken from
the Basis Set Exchange.\cite{Pritchard2019_JCIM_4814}

\begin{table*}
\renewcommand{\arraystretch}{1.}
\caption{
  List of investigated functionals, including the publication year,
  the \textsc{Libxc} identifier, the calculated MSEs for the $1s$,
  $2p$, and $3d$ states as well as the respective OE. Tables
  containing functional rankings by error for the individual states as
  well as the OE can be found in the supplementary material (Tables
  S1--S4). The Libxc identifiers contain information about the
  functional; in addition to the rung of Jacob's ladder: LDA, GGA, or
  meta-GGA (mGGA), hybrid (hyb) functionals are also identifiable from
  the list. \label{tab:xc_funcs} }
    
 \begin{tabular}{lclllll}
 Name & Year & \textsc{Libxc} identifier & \multicolumn{3}{c}{MSE} & OE\ \\ 
      &      &      & 1$s$ & 2$p$ & 3$d$ \ \\ 
$\omega$B97M-V\cite{Mardirossian2016_214110} & 2016 & \texttt{HYB\_MGGA\_XC\_WB97M\_V} & $8.106\plh10^{-4}$ & $5.987\plh10^{-3}$ & $1.128\plh10^{-2}$ & $6.027\plh10^{-3}$ \ \\
$\omega$B97X-D\cite{Chai2008_6615} & 2008 & \texttt{HYB\_GGA\_XC\_WB97X\_D} & $7.641\plh10^{-4}$ & $1.385\plh10^{-2}$ & $2.512\plh10^{-2}$ & $1.324\plh10^{-2}$ \ \\
B3LYP\cite{Becke1988_3098,Lee1988_785,Miehlich1989_200,Becke1993_1372,Stephens1994_11623} & 1994 & \texttt{HYB\_GGA\_XC\_B3LYP} & $7.203\plh10^{-4}$ & $1.055\plh10^{-2}$ & $2.464\plh10^{-2}$ & $1.197\plh10^{-2}$ \ \\
B97-1\cite{Hamprecht1998_6264} & 1998 & \texttt{HYB\_GGA\_XC\_B97\_1} & $3.756\plh10^{-4}$ & $1.108\plh10^{-2}$ & $2.436\plh10^{-2}$ & $1.194\plh10^{-2}$ \ \\
B97M-V\cite{Mardirossian2015_074111} & 2015 & \texttt{MGGA\_XC\_B97M\_V} & $2.466\plh10^{-4}$ & $5.258\plh10^{-3}$ & $1.491\plh10^{-2}$ & $6.804\plh10^{-3}$ \ \\
BHandH\cite{Becke1993_1372} & 1993 & \texttt{HYB\_GGA\_XC\_BHANDH} & $5.122\plh10^{-3}$ & $2.822\plh10^{-3}$ & $4.341\plh10^{-4}$ & $2.793\plh10^{-3}$ \ \\
BLOC\cite{Constantin2013_2256,Constantin2012_035130} & 2013 & \texttt{MGGA\_X\_BLOC,MGGA\_C\_REVTPSS} & $2.046\plh10^{-5}$ & $1.094\plh10^{-2}$ & $2.320\plh10^{-2}$ & $1.139\plh10^{-2}$ \ \\
BLYP\cite{Becke1988_3098,Lee1988_785,Miehlich1989_200} & 1988 & \texttt{GGA\_X\_B88,GGA\_C\_LYP} & $5.790\plh10^{-4}$ & $1.209\plh10^{-2}$ & $2.857\plh10^{-2}$ & $1.374\plh10^{-2}$ \ \\
BLYP35\cite{Renz2009_16292,Kaupp2011_16973} & 2011 & \texttt{HYB\_GGA\_XC\_BLYP35} & $3.902\plh10^{-4}$ & $7.824\plh10^{-3}$ & $1.837\plh10^{-2}$ & $8.861\plh10^{-3}$ \ \\
BOP\cite{Becke1988_3098,Tsuneda1999_10664} & 1999 & \texttt{GGA\_X\_B88,GGA\_C\_OP\_B88} & $5.789\plh10^{-4}$ & $1.209\plh10^{-2}$ & $2.857\plh10^{-2}$ & $1.374\plh10^{-2}$ \ \\
CAM-B3LYP\cite{Yanai2004_51} & 2004 & \texttt{HYB\_GGA\_XC\_CAM\_B3LYP} & $8.286\plh10^{-4}$ & $7.569\plh10^{-3}$ & $1.617\plh10^{-2}$ & $8.188\plh10^{-3}$ \ \\
CAM-QTP00\cite{Verma2014_18A534} & 2014 & \texttt{HYB\_GGA\_XC\_CAM\_QTP\_00} & $4.572\plh10^{-4}$ & $4.187\plh10^{-3}$ & $8.300\plh10^{-3}$ & $4.315\plh10^{-3}$ \ \\
CAM-QTP01\cite{Jin2016_034107} & 2016 & \texttt{HYB\_GGA\_XC\_CAM\_QTP\_01} & $1.339\plh10^{-3}$ & $4.961\plh10^{-3}$ & $9.542\plh10^{-3}$ & $5.281\plh10^{-3}$ \ \\
CAM-QTP02\cite{Haiduke2018_184106} & 2018 & \texttt{HYB\_GGA\_XC\_CAM\_QTP\_02} & $1.714\plh10^{-3}$ & $3.302\plh10^{-3}$ & $6.680\plh10^{-3}$ & $3.899\plh10^{-3}$ \ \\
CHACHIYO\cite{Dirac1930_376,Chachiyo2016_021101} & 2015 & \texttt{LDA\_X,LDA\_C\_CHACHIYO} & $7.711\plh10^{-3}$ & $3.476\plh10^{-3}$ & $1.155\plh10^{-2}$ & $7.579\plh10^{-3}$ \ \\
DM21\cite{Kirkpatrick2021_1385} & 2021 & Uses \PySCF{} implementation instead of \textsc{Libxc} & $2.126\plh10^{-3}$ & $5.435\plh10^{-3}$ & $1.226\plh10^{-2}$ & $6.606\plh10^{-3}$ \ \\
GAM\cite{Yu2015_12146} & 2015 & \texttt{GGA\_X\_GAM,GGA\_C\_GAM} & $2.584\plh10^{-3}$ & $1.561\plh10^{-2}$ & $3.650\plh10^{-2}$ & $1.823\plh10^{-2}$ \ \\
HCTH-93\cite{Hamprecht1998_6264} & 1998 & \texttt{GGA\_XC\_HCTH\_93} & $9.252\plh10^{-4}$ & $1.642\plh10^{-2}$ & $3.689\plh10^{-2}$ & $1.808\plh10^{-2}$ \ \\
HSE03\cite{Heyd2003_8207,Heyd2006_219906} & 2003 & \texttt{HYB\_GGA\_XC\_HSE03} & $1.116\plh10^{-3}$ & $1.150\plh10^{-2}$ & $2.453\plh10^{-2}$ & $1.238\plh10^{-2}$ \ \\
HSE06\cite{Heyd2003_8207,Heyd2006_219906,Krukau2006_224106} & 2006 & \texttt{HYB\_GGA\_XC\_HSE06} & $6.659\plh10^{-4}$ & $8.785\plh10^{-3}$ & $2.052\plh10^{-2}$ & $9.989\plh10^{-3}$ \ \\
HSE12\cite{Moussa2012_204117} & 2012 & \texttt{HYB\_GGA\_XC\_HSE12} & $6.187\plh10^{-4}$ & $8.201\plh10^{-3}$ & $1.911\plh10^{-2}$ & $9.309\plh10^{-3}$ \ \\
LC-QTP\cite{Haiduke2018_184106} & 2018 & \texttt{HYB\_GGA\_XC\_LC\_QTP} & $2.276\plh10^{-3}$ & $3.818\plh10^{-3}$ & $7.590\plh10^{-3}$ & $4.561\plh10^{-3}$ \ \\
LC-VV10\cite{Vydrov2010_244103} & 2010 & \texttt{HYB\_GGA\_XC\_LC\_VV10} & $1.089\plh10^{-3}$ & $6.905\plh10^{-3}$ & $1.166\plh10^{-2}$ & $6.553\plh10^{-3}$ \ \\
LRC-$\omega$PBE\cite{Rohrdanz2009_054112} & 2009 & \texttt{HYB\_GGA\_XC\_LRC\_WPBE} & $1.121\plh10^{-3}$ & $9.294\plh10^{-3}$ & $1.668\plh10^{-2}$ & $9.033\plh10^{-3}$ \ \\
M06-L\cite{Zhao2006_194101,Zhao2008_215} & 2006 & \texttt{MGGA\_X\_M06\_L,MGGA\_C\_M06\_L} & $9.242\plh10^{-4}$ & $1.521\plh10^{-2}$ & $3.368\plh10^{-2}$ & $1.660\plh10^{-2}$ \ \\
M11-L\cite{Peverati2012_117} & 2012 & \texttt{MGGA\_X\_M11\_L,MGGA\_C\_M11\_L} & $2.320\plh10^{-3}$ & $1.876\plh10^{-2}$ & $4.593\plh10^{-2}$ & $2.234\plh10^{-2}$ \ \\
MN12-L\cite{Peverati2012_13171} & 2012 & \texttt{MGGA\_X\_MN12\_L,MGGA\_C\_MN12\_L} & $1.541\plh10^{-3}$ & $5.796\plh10^{-3}$ & $2.036\plh10^{-2}$ & $9.234\plh10^{-3}$ \ \\
MN15\cite{Yu2016_5032} & 2016 & \texttt{HYB\_MGGA\_X\_MN15,MGGA\_C\_MN15} & $2.532\plh10^{-4}$ & $9.867\plh10^{-3}$ & $2.133\plh10^{-2}$ & $1.048\plh10^{-2}$ \ \\
MN15-L\cite{Yu2016_1280} & 2016 & \texttt{MGGA\_X\_MN15\_L,MGGA\_C\_MN15\_L} & $2.118\plh10^{-3}$ & $4.673\plh10^{-3}$ & $1.327\plh10^{-2}$ & $6.689\plh10^{-3}$ \ \\
MS0\cite{Sun2012_051101,Perdew2009_026403,Perdew2011_179902} & 2012 & \texttt{MGGA\_X\_MS0,GGA\_C\_REGTPSS} & $5.973\plh10^{-4}$ & $1.121\plh10^{-2}$ & $2.233\plh10^{-2}$ & $1.138\plh10^{-2}$ \ \\
MS1\cite{Sun2013_044113,Perdew2009_026403,Perdew2011_179902} & 2013 & \texttt{MGGA\_X\_MS1,GGA\_C\_REGTPSS} & $6.013\plh10^{-4}$ & $1.161\plh10^{-2}$ & $2.346\plh10^{-2}$ & $1.189\plh10^{-2}$ \ \\
MS2\cite{Sun2013_044113,Perdew2009_026403,Perdew2011_179902} & 2013 & \texttt{MGGA\_X\_MS2,GGA\_C\_REGTPSS} & $6.039\plh10^{-4}$ & $1.193\plh10^{-2}$ & $2.432\plh10^{-2}$ & $1.229\plh10^{-2}$ \ \\
OLYP\cite{Handy2001_403,Lee1988_785} & 2009 & \texttt{GGA\_X\_OPTX,GGA\_C\_LYP} & $3.934\plh10^{-4}$ & $1.309\plh10^{-2}$ & $2.971\plh10^{-2}$ & $1.440\plh10^{-2}$ \ \\
PBE\cite{Perdew1996_3865,Perdew1997_1396} & 1996 & \texttt{GGA\_X\_PBE,GGA\_C\_PBE} & $9.196\plh10^{-4}$ & $1.096\plh10^{-2}$ & $2.538\plh10^{-2}$ & $1.242\plh10^{-2}$ \ \\
PBEsol\cite{Perdew2008_136406} & 2007 & \texttt{GGA\_X\_PBE\_SOL,GGA\_C\_PBE\_SOL} & $3.799\plh10^{-3}$ & $6.918\plh10^{-3}$ & $2.010\plh10^{-2}$ & $1.027\plh10^{-2}$ \ \\
PKZB\cite{Perdew1999_2544} & 1999 & \texttt{MGGA\_X\_PKZB,MGGA\_C\_PKZB} & $9.408\plh10^{-4}$ & $1.038\plh10^{-2}$ & $2.470\plh10^{-2}$ & $1.201\plh10^{-2}$ \ \\
PW91\cite{Ziesche1991_,Perdew1992_6671,Perdew1993_4978} & 1992 & \texttt{GGA\_X\_PW91,GGA\_C\_PW91} & $7.771\plh10^{-4}$ & $1.086\plh10^{-2}$ & $2.449\plh10^{-2}$ & $1.204\plh10^{-2}$ \ \\
QTP17\cite{Jin2018_064111} & 2018 & \texttt{HYB\_GGA\_XC\_QTP17} & $3.060\plh10^{-3}$ & $1.483\plh10^{-3}$ & $4.612\plh10^{-3}$ & $3.051\plh10^{-3}$ \ \\
RPBE\cite{Hammer1999_7413,Perdew1996_3865,Perdew1997_1396} & 1999 & \texttt{GGA\_X\_RPBE,GGA\_C\_PBE} & $4.576\plh10^{-4}$ & $1.346\plh10^{-2}$ & $2.932\plh10^{-2}$ & $1.441\plh10^{-2}$ \ \\
SPW92\cite{Dirac1930_376,Perdew1992_13244} & 1992 & \texttt{LDA\_X,LDA\_C\_PW\_MOD} & $7.702\plh10^{-3}$ & $3.392\plh10^{-3}$ & $1.138\plh10^{-2}$ & $7.490\plh10^{-3}$ \ \\
SVWN\cite{Dirac1930_376,Bloch1929_545,Vosko1980_1200} & 1980 & \texttt{LDA\_X,LDA\_C\_VWN} & $7.707\plh10^{-3}$ & $3.390\plh10^{-3}$ & $1.138\plh10^{-2}$ & $7.492\plh10^{-3}$ \ \\
TASK\cite{Aschebrock2019_033082,Perdew1992_13244} & 2019 & \texttt{MGGA\_X\_TASK,LDA\_C\_PW} & $2.523\plh10^{-3}$ & $1.463\plh10^{-2}$ & $2.698\plh10^{-2}$ & $1.471\plh10^{-2}$ \ \\
TM\cite{Tao2016_073001} & 2016 & \texttt{MGGA\_X\_TM,MGGA\_C\_TM} & $2.972\plh10^{-5}$ & $1.035\plh10^{-2}$ & $2.240\plh10^{-2}$ & $1.093\plh10^{-2}$ \ \\
TPSS\cite{Tao2003_146401,Perdew2004_6898} & 2003 & \texttt{MGGA\_X\_TPSS,MGGA\_C\_TPSS} & $2.046\plh10^{-5}$ & $1.094\plh10^{-2}$ & $2.318\plh10^{-2}$ & $1.138\plh10^{-2}$ \ \\
TPSSh\cite{Staroverov2003_12129} & 2003 & \texttt{HYB\_MGGA\_XC\_TPSSH} & $1.640\plh10^{-5}$ & $9.835\plh10^{-3}$ & $2.080\plh10^{-2}$ & $1.022\plh10^{-2}$ \ \\
XLYP\cite{Xu2004_2673} & 2004 & \texttt{GGA\_XC\_XLYP} & $1.369\plh10^{-4}$ & $1.217\plh10^{-2}$ & $2.792\plh10^{-2}$ & $1.341\plh10^{-2}$ \ \\
r$^{2}$SCAN\cite{Furness2020_8208,Furness2020_9248} & 2020 & \texttt{MGGA\_X\_R2SCAN,MGGA\_C\_R2SCAN} & $1.495\plh10^{-5}$ & $8.087\plh10^{-3}$ & $1.617\plh10^{-2}$ & $8.091\plh10^{-3}$ \ \\
r$^{2}$SCAN0\cite{Bursch2022_JCP_134105} & 2022 & Custom-defined in \PySCF{} & $8.173\plh10^{-6}$ & $6.051\plh10^{-3}$ & $1.204\plh10^{-2}$ & $6.035\plh10^{-3}$ \ \\
r$^{2}$SCAN50\cite{Bursch2022_JCP_134105} & 2022 & Custom-defined in \PySCF{} & $3.504\plh10^{-6}$ & $4.025\plh10^{-3}$ & $7.975\plh10^{-3}$ & $4.001\plh10^{-3}$ \ \\
r$^{2}$SCANh\cite{Bursch2022_JCP_134105} & 2022 & Custom-defined in \PySCF{} & $1.198\plh10^{-5}$ & $7.271\plh10^{-3}$ & $1.451\plh10^{-2}$ & $7.266\plh10^{-3}$ \ \\
rSCAN\cite{Bartok2019_161101} & 2019 & \texttt{MGGA\_X\_RSCAN,MGGA\_C\_RSCAN} & $1.495\plh10^{-5}$ & $8.087\plh10^{-3}$ & $1.617\plh10^{-2}$ & $8.091\plh10^{-3}$ \ \\
rSCAN0\cite{Bursch2022_JCP_134105} & 2022 & Custom-defined in \PySCF{} & $8.173\plh10^{-6}$ & $6.051\plh10^{-3}$ & $1.204\plh10^{-2}$ & $6.034\plh10^{-3}$ \ \\
rSCAN50\cite{Bursch2022_JCP_134105} & 2022 & Custom-defined in \PySCF{} & $3.504\plh10^{-6}$ & $4.025\plh10^{-3}$ & $7.975\plh10^{-3}$ & $4.001\plh10^{-3}$ \ \\
rSCANh\cite{Bursch2022_JCP_134105} & 2022 & Custom-defined in \PySCF{} & $1.198\plh10^{-5}$ & $7.271\plh10^{-3}$ & $1.451\plh10^{-2}$ & $7.266\plh10^{-3}$ \ \\
revPBE\cite{Perdew1996_3865,Perdew1997_1396,Zhang1998_890} & 1998 & \texttt{GGA\_X\_PBE\_R,GGA\_C\_PBE} & $4.572\plh10^{-4}$ & $1.358\plh10^{-2}$ & $2.995\plh10^{-2}$ & $1.466\plh10^{-2}$ \ \\
revTPSS\cite{Perdew2009_026403,Perdew2011_179902} & 2009 & \texttt{MGGA\_X\_REVTPSS,MGGA\_C\_REVTPSS} & $1.302\plh10^{-5}$ & $1.040\plh10^{-2}$ & $2.186\plh10^{-2}$ & $1.076\plh10^{-2}$ \ \\
\end{tabular}
\end{table*}

The \textsc{Libxc} library\cite{Lehtola2018_S_1}---which implements
over 600 DFAs---is used in \PySCF{} to evaluate the DFAs. The library
provides access to a vast variety of DFAs, of which 49 were chosen for
this work; see \tabref{xc_funcs} for the complete list of investigated
functionals. Our selection includes functionals of the first to the
fourth rung of Jacob's ladder,\cite{Perdew2001_ACP_1} that is, local
density approximations (LDAs), generalized gradient approximations
(GGAs), meta-GGAs, as well as global and range-separated hybrid
functionals. In addition, we consider six hybrids of rSCAN and
r$^2$SCAN with varying fractions of Hartree--Fock exchange discussed
in \citeref{Bursch2022_JCP_134105}; these functionals were defined in
the \PySCF{} input files as weighted combinations of r($^2$)SCAN
exchange and Hartree--Fock exchange + 100\% r($^2$)SCAN
correlation. The DM21 functional was also chosen for this study; we
use the original implementation in \PySCF{}
of \citet{Kirkpatrick2021_1385}. This brings up the total to 56
functionals for this study.  An unpruned (300,590) quadrature grid is
used in all calculations, including the non-local correlation
component in B97M-V, $\omega$B97M-V and LC-VV10.

As the total energies scale as $E_n \propto Z^2$ according
to \eqref{Eexact}, the results will be analyzed in terms of absolute
relative errors (AREs). Hydrogenic estimates show that the
approximated exchange-correlation energy scales like $Z$ in the large
$Z$ limit,\cite{Kaplan2020_JCP_74114} meaning that the relative errors
should tend to zero like $1/Z$. The ARE for a given state of a given
ion is given by
\begin{equation}
    \text{ARE} =  |(E_{\text{calc}}-E_{\text{ref}})/E_{\text{ref}}|.
\end{equation}
The information in the AREs is analyzed with two further error
metrics. The mean state error (MSE) measures the overall functional
error over all ions by averaging the ARE over all ions
\begin{equation}
    \text{MSE} = 1/N_{\text{ions}} \sum_{i}^{N_{\text{ions}}} \text{ARE}_{i}.
\end{equation}
The overall error (OE) for a functional is obtained by further
averaging the MSE over all considered states ($1s$, $2p$, and $3d$)
\begin{equation}
    \text{OE} = 1/N_{\text{states}} \sum_{i}^{N_{\text{states}}} \text{MSE}_{i}. 
\end{equation}

\section{Results \label{sec:results}}
\subsection{Basis set convergence \label{sec:basconv}}
Before pursuing density functional calculations, we analyze the basis
set truncation errors (BSTEs) for the one-electron cations in the
polarization consistent and hydrogenic Gaussian basis sets.  We aim
for a mean BSTE smaller than $5 \times 10^{-5}$ $E_{\text{h}}$ for the
whole benchmark set ranging from \ce{H^{0}} to \ce{Kr^{35+}} to ensure
that our results are converged close to the complete basis set limit.

Unrestricted Hartree--Fock (UHF) is exact for one-electron
systems and thereby gives the exact energy $E_{n}^\text{UHF}$ in the
studied basis; the difference of $E_{n}^\text{UHF/basis}$ and the exact
analytical energy (\eqref{Eexact})
\begin{equation}
  \Delta^\text{basis}_{n} = E_{n}^\text{UHF/basis} - E_{n} \geq 0 \label{eq:bste}
\end{equation}
is therefore a variational measure of the BSTE for the state with
given $n$ of the studied hydrogenic ions.

\begin{figure}
    \centering \includegraphics[width=\linewidth]{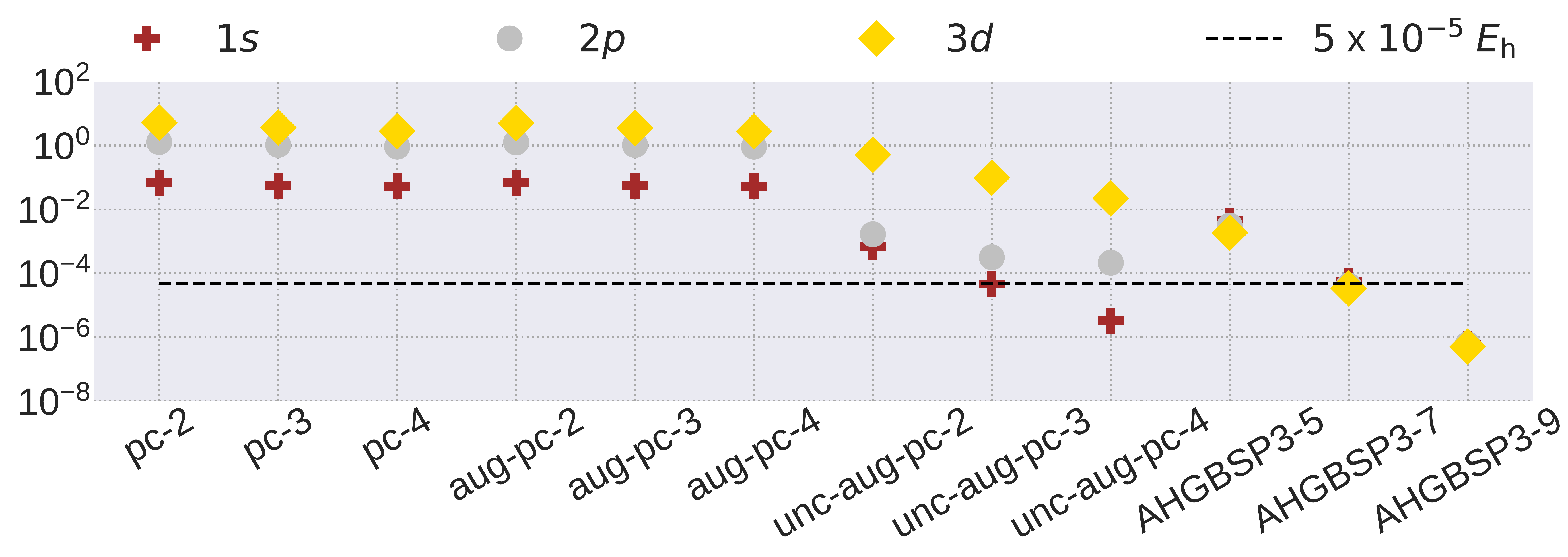} \caption{Mean
    basis set truncation error (ME) in $E_{\text{h}}$ at UHF level of theory for the
    $1s$ ground state and the $2p$ and $3d$ excited states,
    respectively. The reference values are calculated
    with \eqref{Eexact}, and the aimed accuracy threshold $5 \times
    10^{-5} E_{\text{h}}$ is shown with the dashed horizontal
    line.} \label{fig:mbste_oee}
\end{figure}

The calculated mean BSTEs for a variety of polarization consistent and
hydrogenic Gaussian basis sets are shown in \figref{mbste_oee};
additional results can be found in the supplementary
material. Unsurprisingly, uncontracting the \mbox{(aug-)pc-$n$} basis
sets---yielding the
\mbox{unc-(aug-)pc-$n$} basis sets---results in a noticeable decrease of the
BSTE, because the contractions were determined in
\citeref{Jensen2001_JCP_9113} with the BLYP functional\cite{Becke1988_3098,
Lee1988_785, Miehlich1989_200} that suffers from SIE for the
$1s$ state, while the $p$ and $d$ functions in the basis set describe
either polarization effects or the occupied $p$ or $d$ orbitals in the
screened neutral atom. Although the large uncontracted polarization
consistent basis sets exhibit satisfactory performance for the $1s$
state, they result in much larger errors for the $2p$ and $3d$ states;
this error is again caused by the $p$ and $d$ orbitals in the neutral
atom being screened by the core electrons, which results in the lack
of tight $p$ and $d$ basis functions that are necessary for the $2p$
and $3d$ states of the one-electron ions.

In contrast, the primitive (not contracted) hydrogenic Gaussian basis
sets of \citeref{Lehtola2020_JCP_134108} show uniform accuracy for the
$1s$, $2p$, and $3d$ states, and as can be observed in
\figref{mbste_oee}, the targeted mean BSTE threshold is roughly
achieved already with the
\mbox{AHGBSP3-7} basis set. The \mbox{AHGBSP3-9} basis sets yield
errors below the desired threshold for all states, and is therefore
chosen for all the remaining calculations of this study.

Although this analysis was limited to Hartree--Fock calculations, we
note that the basis set requirements of Hartree--Fock and DFT are
known to be similar.\cite{Christensen2000_CPL_400} Furthermore,
reliable reference energies for DFAs can be obtained with fully
numerical methods,\cite{Lehtola2019_IJQC_25968,
Lehtola2019_IJQC_25945, Lehtola2020_PRA_12516} and exploratory
calculations presented in the supplementary material confirm that the
BSTEs in the AHGBSP3-9 basis are small also for
DFAs.

\subsection{OEE cation benchmark \label{sec:oeebenchmark}}

\subsubsection{Exploratory analysis \label{sec:graphics}}
\begin{figure}
    \centering
    \subfigure[$1s$]{\includegraphics[width=\linewidth]{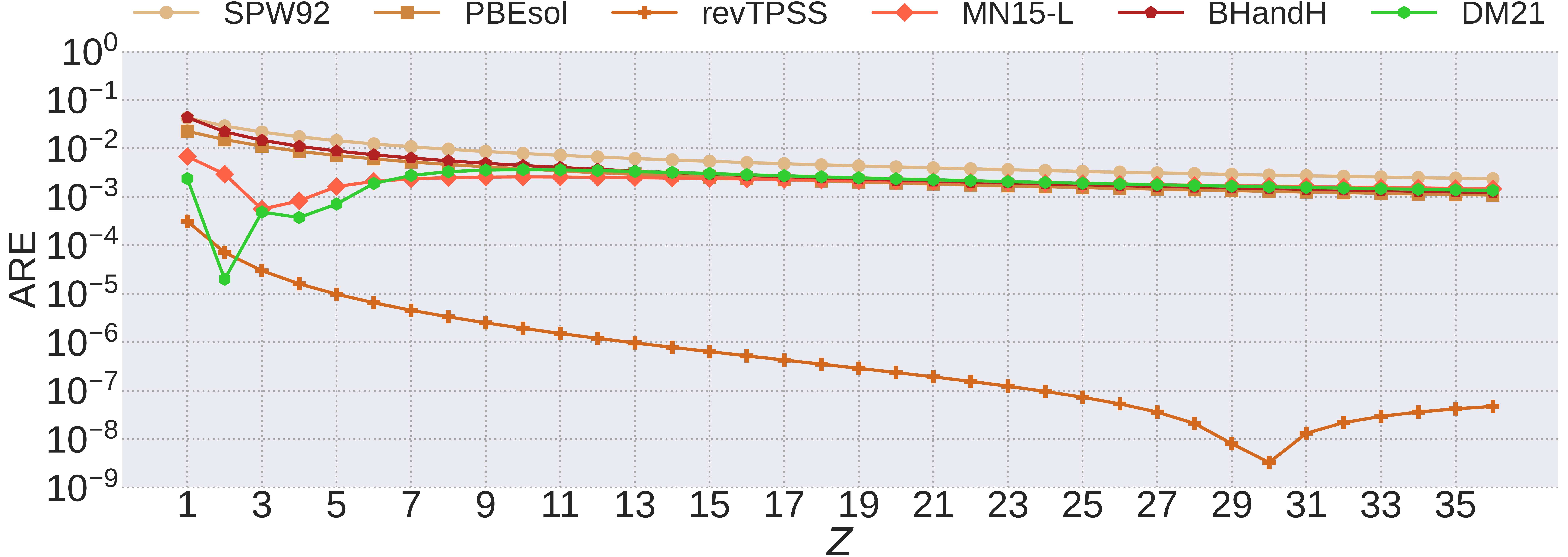} \label{fig:oee1s}}
    \subfigure[$2p$]{\includegraphics[width=\linewidth]{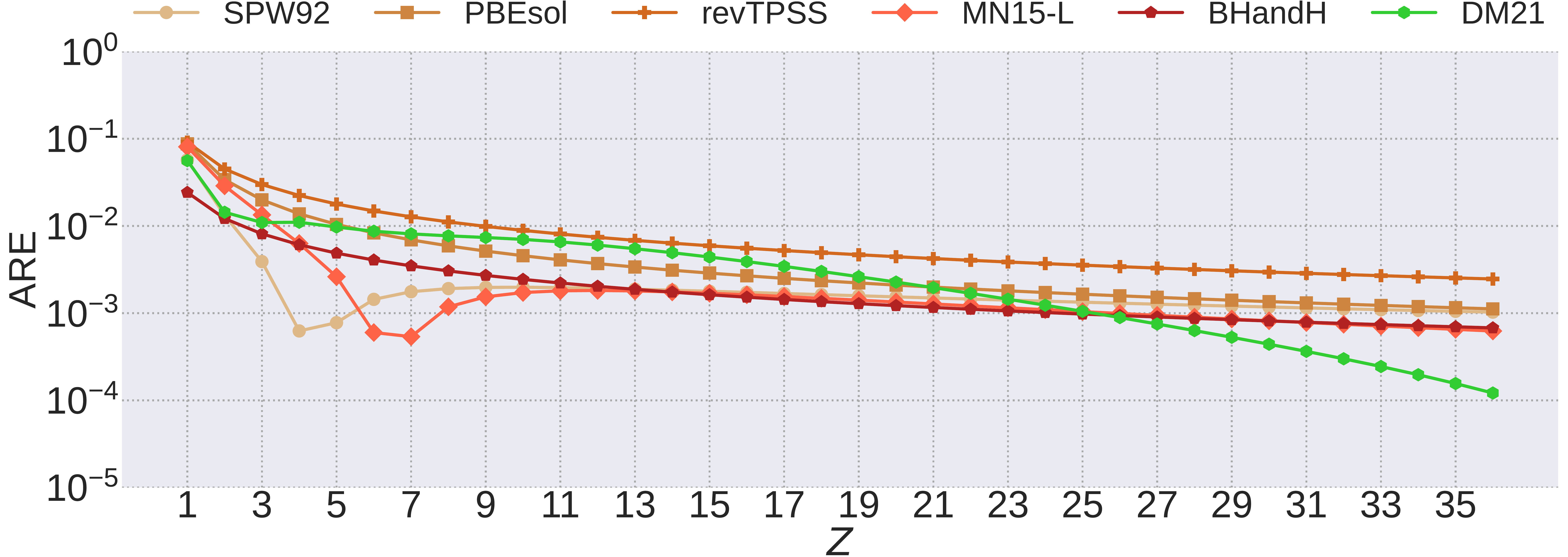} \label{fig:oee2p}}
    \subfigure[$3d$]{\includegraphics[width=\linewidth]{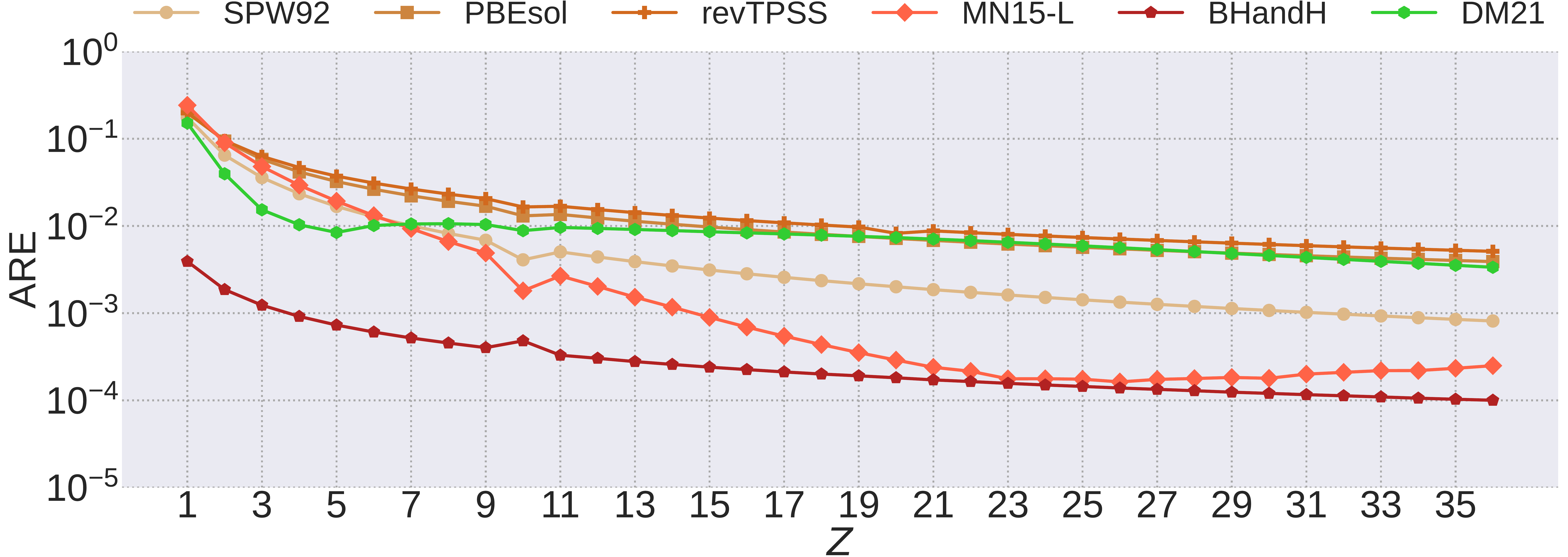} \label{fig:oee3d}}

    \caption{Functional errors for the $1s$, $2p$, and $3d$ states for
    the SPW92, PBEsol, revTPSS, MN15-L, BHandH, and DM21
    functionals. \label{fig:oee_species}}
\end{figure}

We begin the analysis by a graphical study of the results of the
SPW92, PBEsol, revTPSS, MN15-L, BHandH and DM21 functionals
in \figref{oee_species}. This limited set of functionals contains LDA,
GGA, and meta-GGA functionals from first principles (SPW92, PBEsol,
and revTPSS, respectively), semiempirical functionals (MN15-L and
DM21) as well as hybrid functionals (BHandH and DM21).

As will be discussed in \secref{fullanal}, revTPSS is the most
accurate meta-GGA functional for the $1s$
state. In \figref{oee_species}, revTPSS is outperformed by DM21 only
for \ce{He+}, and otherwise revTPSS affords much lower errors than the
five other functionals in the figure. In contrast, the performance of
DM21 is inconsistent. DM21 has lower errors for light ions than for
heavy ions, but the curve is kinked for the light ions. DM21's curve
becomes smooth for heavy ions, but DM21 is also less accurate for
heavy ions. MN15-L also shows a kinky behavior with lower errors for
light ions; these non-systematic features of DM21 and MN15-L can be
tentatively explained by their semiempirical character; the curves for
the first principles functionals are smoother.

The functional errors for the $2p$ state are shown in \figref{oee2p}.
The performance for the $2p$ state is strikingly different compared to
the $1s$ state shown in \figref{oee1s}. The plots for the $2p$ state
in \figref{oee2p} show more structure and curve crossings. The
behavior of DM21 is qualitatively different from that of the other
functionals: DM21 shows large relative errors for light atoms and
lower relative errors for heavy atoms, while most of the other
functionals shown behave similarly to each other. The only other
exceptions to this are the SPW92 and MN15-L functionals that show dips
at $Z \simeq 3$ and $Z \simeq 4$, respectively; the two functionals
are thus oddly more accurate for some values of $Z$ than others.

The errors for the $3d$ state are shown in \figref{oee3d}. BHandH has
small errors for all ions for the $3d$ state. The behavior of DM21 and
MN15-L again differs qualitatively from the other functionals. While
DM21 shows less variation for the $3d$ state than for the $2p$ state,
MN15-L does the opposite: MN15-L has large errors for light ions,
becomes nearly as accurate as BHandH for $Z \simeq 22$, while the
relative error increases again for heavier ions. 


\subsubsection{Full analysis \label{sec:fullanal}}

The MSEs and OE for all studied functionals are shown
in \tabref{xc_funcs}. Although \tabref{xc_funcs} contains all of the
data used in the present analysis, additional tables showing the
rankings of the functionals in terms of the errors for the $1s$, $2p$
and $3d$ states as well as in terms of the overall error can be found
in the supplementary material.

Clearly, the performance of all LDAs is practically identical. This
suggests that the functional error for LDAs is limited by the simple
functional form. While LDAs show larger errors than GGAs and meta-GGAs
for the $1s$ state, they perform better than GGAs and meta-GGAs for
the $2p$ and $3d$ states.

PBEsol\cite{Perdew2008_136406} is the GGA that yields the smallest
errors for the $2p$ and $3d$ states, as well as the smallest overall
error. The XLYP GGA has a lower error than PBEsol for the $1s$
state. Although XLYP and even PBEsol are better for $1s$ states than
any LDA, they have higher OEs than any LDA because of their
considerably poorer performance for the $2p$ and $3d$
states. Analogous findings apply also to all other studied GGAs.

The best meta-GGA for the $1s$ state is
revTPSS,\cite{Perdew2009_026403, Perdew2011_179902} closely followed
by rSCAN,\cite{Bartok2019_161101} and
r$^{2}$SCAN\cite{Furness2020_8208} (see \tabref{xc_funcs} or the
supplementary material). The best meta-GGA in terms of overall error
is MN15-L.\cite{Yu2016_1280}

Hybrid functionals have better accuracy, as they contain some
Hartree--Fock exchange which is free of self-interaction. The best
hybrid GGA functionals in terms of overall error are
BHandH\cite{Becke1993_1372} and QTP17.\cite{Jin2018_064111} BHandH has
low MSEs for all states and has the best overall performance, which
can be understood by its composition of 50\% of Hartree--Fock exchange
and 50\% LDA exchange + 100\% Lee--Yang--Parr correlation.  QTP17 has
the second best performance for all states; it, too, contains a
mixture of Hartree--Fock (62\%) and LDA exchange (38\%).

\begin{figure}
  \centering
  \includegraphics[width=\linewidth]{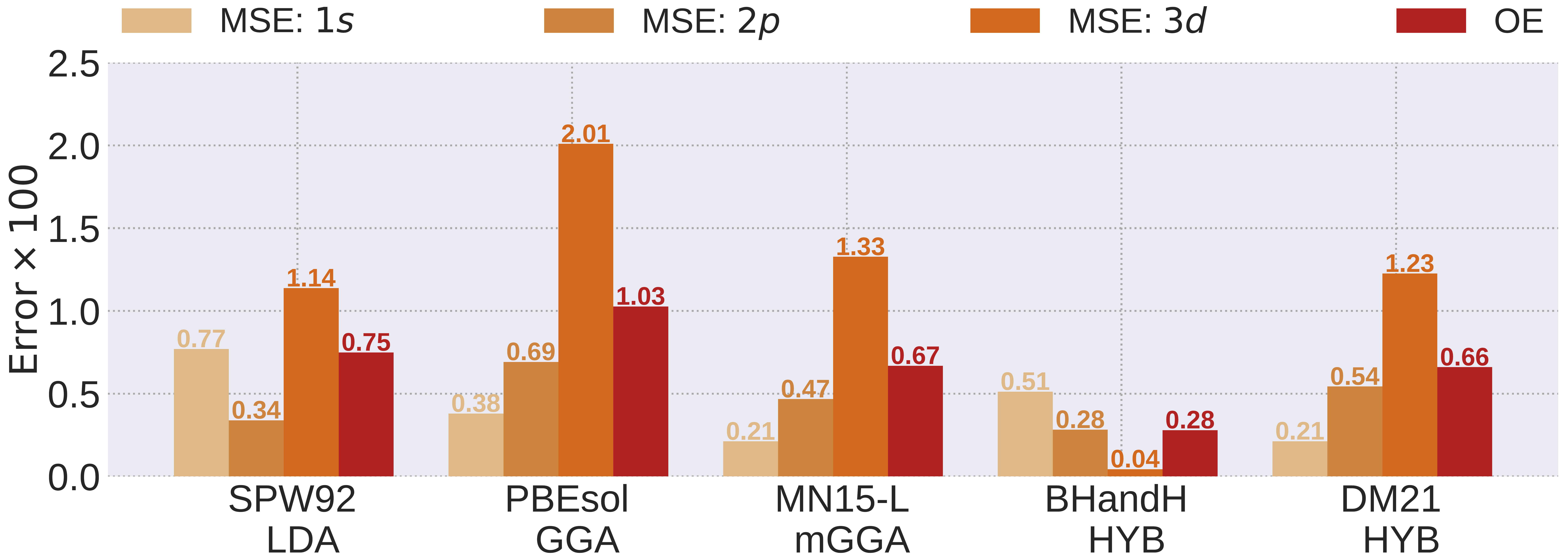}
  \caption{
    MSEs and OEs for the best functional of each rung of Jacob's
    ladder of all the investigated functionals.
  }
  \label{fig:oee_barchart}
\end{figure}

The best functionals of each rung in terms of overall error are SPW92,
PBEsol, MN15-L, and BHandH, respectively. The corresponding error
distributions are summarized in comparison to DM21
in \figref{oee_barchart}.  Interestingly, the performance of the DM21
functional appears similar to that of MN15-L.

\begin{figure*}
    \centering
    \includegraphics[width=\linewidth]{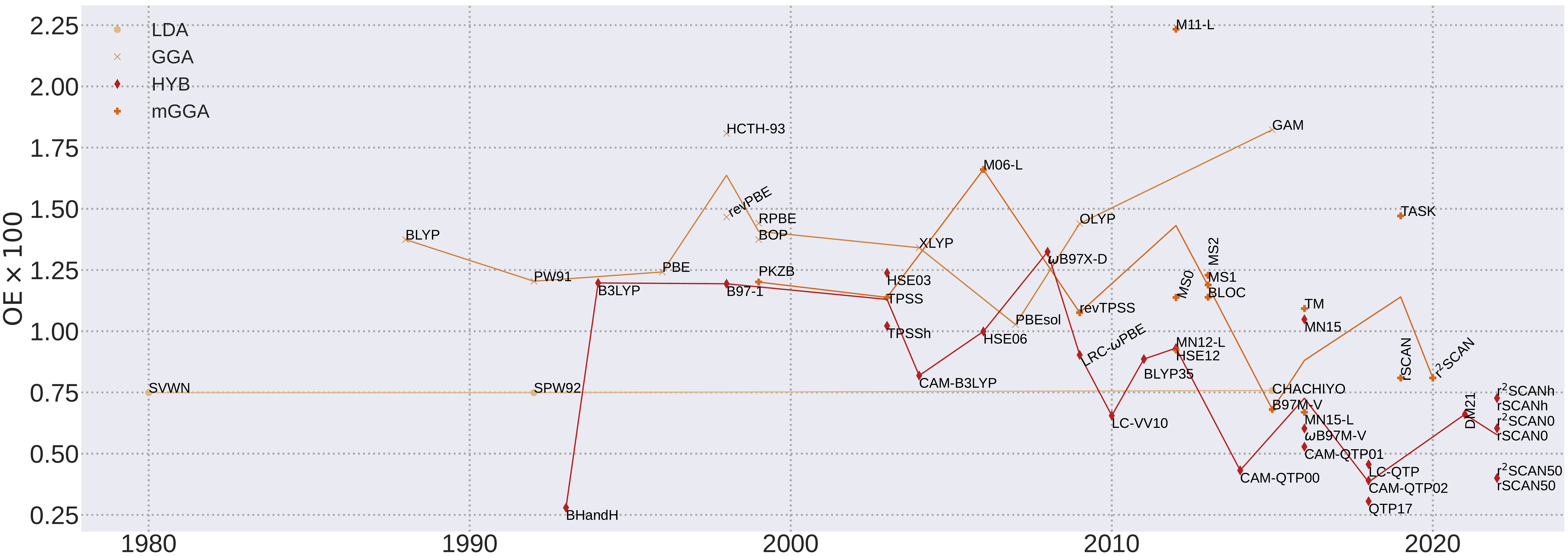}
    \caption{
      The overall functional error (OE) for various types of
      functionals, plotted as a function of the publication
      year. Yearly averages are shown as solid lines as a guide to the
      eye.}
    \label{fig:oee_scatter}
\end{figure*}

Following \citet{Medvedev2017_S_49}, the calculated OE for all
functionals and ions is plotted against the publication year
in \figref{oee_scatter}. As is clear from this plot, the improvement
in one-electron error is not fully systematic and features a
significant amount of spread and some notable outliers like M11-L,
GAM, and TASK. However, in the recent decade, hybrid functionals have
become better overall. As an example, the various QTP functionals
dominate the bottom right of the figure; these functionals are closely
related in functional form and contain high amounts of Hartree--Fock
exchange which decreases the one-electron error. Unsurprisingly,
hybrid functionals based on rSCAN and r$^{2}$SCAN perform well, and
the functionals with large fractions of Hartree--Fock exchange share
the bottom right of the figure with the QTP functionals.

\subsubsection{Comparison to literature data \label{sec:litcomp}}

The study of \citet{Lonsdale2020_PCCP_15805} employed an uncontracted
aug-cc-pVQZ basis set\cite{Dunning1989_JCP_1007, Kendall1992_JCP_6796}
with the following nucleus dependent quadrature grids for the study of
$1s$ states of hydrogenic cations: (45,770) for H and He, (50, 770)
for Li--Ne, (55, 770) for Na--Ar and (60, 770) for K--Kr. However, it
appears that K and Ca were excluded
from \citeref{Lonsdale2020_PCCP_15805} (see caption of Figs. 3 and 10
in \citeref{Lonsdale2020_PCCP_15805}).

Our trends and absolute values for the MSE for the $1s$ state are in
satisfactory agreement for the subset of functionals studied in both
works, although we did identify basis set incompleteness issues in
some results of \citeref{Lonsdale2020_PCCP_15805} as discussed in the
supplementary material. The largest basis set incompleteness effects
are observed for the M11-L and M06-L Minnesota functionals, which are
known to converge remarkably slowly to the basis set
limit.\cite{Mardirossian2013_JCTC_4453}
 
\citet{Lonsdale2020_PCCP_15805} only studied one LDA functional (SVWN);
we considered more LDAs and found them to have similar
performance. \citet{Lonsdale2020_PCCP_15805} found OLYP to be the best
GGA functional for the $1s$ state; we also considered XLYP and found
it to yield a considerably lower MSE for the $1s$ state than OLYP.
\citet{Lonsdale2020_PCCP_15805} included a broader set of hybrid
functionals separating global, range-separated hybrids and double
hybrids; however, our main motivation is the connection to
self-interaction corrected methods where hybrid functionals are
typically not used.  We found rSCAN50/r$^{2}$SCAN50 to be the best
hybrid functional for the $1s$ state,
while \citet{Lonsdale2020_PCCP_15805} determined TPSSh and SCAN0 to be
the best hybrids. All rSCAN and r$^{2}$SCAN based hybrid functionals,
i.e., rSCANh/r$^{2}$SCANh, rSCAN0/r$^{2}$SCAN0, and
rSCAN50/r$^{2}$SCAN50 as well as TPSSh have a good performance for the
$1s$ state.  The revTPSS functional is found in our work as well
by \citet{Lonsdale2020_PCCP_15805} to be the best non-hybrid meta-GGA
functional for the $1s$ state.

\section{Summary and discussion \label{sec:summary}} 

We used exactly solveable hydrogenic cations in their $1s$ ground
state and $2p$ and $3d$ excited states to determine the
self-consistent one-electron error for 56 density functionals
including the novel DM21 of \citet{Kirkpatrick2021_1385}, employing
the methodology of \citet{Gunnarsson1976_PRB_4274} for the excited
state calculations. In accordance with an earlier finding by
\citet{Sun2016_JCP_191101} apparently based on non-self-consistent
calculations for the hydrogen atom and molecule and one LDA
functional, we find for 36 hydrogenic cations that all LDAs perform
better for the excited $2p$ and $3d$ states than any of the tested
GGAs and meta-GGAs. The performance of various LDAs appears to be
almost identical, as the calculated errors are nearly
indistinguishable, suggesting that the errors are limited by the
simple functional form used in LDAs. \citet{Sun2016_JCP_191101}
pointed out that larger errors for excited states are a necessary
consequence of orbital nodality.

The revTPSS functional is the best performing meta-GGA for the $1s$
state, tightly followed by the rSCAN and r$^2$SCAN functionals. MN15-L
shows a better overall performance than LDAs for all states; however,
the performance of MN15-L is non-systematic like that of DM21.

Hybrid functionals like BHandH and QTP17 have the best overall
performance as they explictly include some fraction of Hartree--Fock
exchange. Moreover, both BHandH and QTP17 are mixtures of
Hartree--Fock and LDA exchange, leading to the good observed accuracy.

DM21 turns out to be only close to exact for the $1s$ state OEE from
\ce{H^{0}} to \ce{B^{4+}} (see \figref{oee_species}).  For \ce{He^{+}} to
\ce{B^{4+}} DM21 shows also good performance for $2p$ and $3d$ states.
However, over the whole range of investigated species \ce{H^{0}} to
\ce{Kr^{35+}}, DM21 exhibits various trend changes and an overall
inconsistent performance.  Thus, one might improve the next generation
of the DM21 functional by including more one-electron cations in the
training sets for various elements in the periodic table.  This might
increase the consistency of promising properties of such kind of
machine-learned functionals.

We found PBEsol to be the most accurate GGA functional for the $2p$
and $3d$ states. PBEsol is also the most accurate GGA functional
overall. These findings are interesting to contrast with that
of \citet{Lehtola2016_JCTC_4296}, who showed that PBEsol is one of the
few functionals whose accuracy improves when PZ-SIC is applied with
complex orbitals. The development of novel DFAs with reduced
one-electron error could therefore be useful for PZ-SIC calculations,
as the reduced one-electron errors (\eqref{sie}) would affect the
numerics of the PZ correction (\eqref{pz}) and might alleviate
well-known issues with PZ-SIC and PZFLO-SIC discussed in
\secref{intro}.

\textbf{Note added in proof}
After the acceptance of this paper, we became aware of a preprint
by \citet{Lonsdale2022_} that includes discussion on excited states of
hydrogenic cations.

\section*{Supplementary material}

Exploratory finite element studies of basis set truncation errors in
density functional calculations. Sorted rankings of the functionals by
errors for the $1s$, $2p$ and $3d$ states as well as the overall
error. Bar plots of the errors for all studied functionals. Comparison
of the $1s$ data to the study of \citet{Lonsdale2020_PCCP_15805} with
additional basis set incompleteness studies. Tables of the calculated
total energies for the $1s$, $2p$, and $3d$ states for all studied
functionals.

\section*{Author Declarations}
The authors have no conflicts to disclose.

\section*{Acknowledgment}

We thank Jens Kortus for valuable comments on the manuscript.
S. Schwalbe has been funded by the Deutsche Forschungsgemeinschaft
(DFG, German Research Foundation) - Project ID 421663657 - KO
1924/9-2.  This work is part of the OpenSIC project.  We thank ZIH
Dresden for computational time and support.

\bibliography{refs.bib,susi.bib,libxc.bib}
\end{document}